\documentclass[10pt,letterpaper,twocolumn]{article} 

\usepackage{ol2}
\usepackage[colorlinks=true,urlcolor=blue,citecolor=blue,linkcolor=blue]{hyperref}
\usepackage{amsmath}

\newcommand{\ket}[1]{|#1\rangle}

\providecommand{\e}[1]{\ensuremath{\times 10^{#1}}}

\begin{document}

\twocolumn[ 

\title{A laser based accelerator for ultracold atoms}


\author{A.~Rakonjac,$^{1}$ A.~B.~Deb,$^1$ S.~Hoinka,$^{1,2}$ D.~Hudson,$^{1,3}$ B.~J.~Sawyer,$^1$ and N.~Kj{\ae}rgaard$^{1,*}$}
\address{
$^1$Jack Dodd Centre for Quantum Technology,
Department of Physics, University of Otago, Dunedin, New Zealand\\
$^2$Currently with the Centre for Atom Optics and Ultrafast Spectroscopy,\\ Swinburne University of Technology, Melbourne 3122, Australia
\\
$^3$Currently with the Centre for Ultrahigh bandwidth Devices for Optical Systems, Institute of Photonics and Optics Science,\\The University of Sydney,
Sydney, NSW 2006, Australia
\\
$^*$Corresponding author: niels.kjaergaard@otago.ac.nz}
\begin{abstract}We present first results on our implementation of a laser based accelerator for ultracold atoms. Atoms cooled to a temperature of 420 nK are confined and accelerated by means of laser tweezer beams and the atomic scattering is directly observed in laser absorption imaging. The optical collider has been characterized using $\rm ^{87}Rb$ atoms in the $\ket{F=2,m_F=2}$ state, but the scheme is not restricted to atoms in any particular magnetic substates and can readily be extended to other atomic species as well.
\end{abstract}
 ] 
\noindent The scattering experiment has been described as the most important experimental technique in quantum physics \cite{Taylor2006}. Typical textbook treatments describe the problem in terms of a beam impinging on a target: particles collide along a \textit{well-defined} axis with respect to which the angular scattering distribution of particles is determined. Knowledge about collision properties of neutral atoms at extremely low energies has been of crucial importance for developing the field of ultracold atomic physics \cite{Weiner1999}. So far, the investigation of cold collisions has predominantly been carried out with atoms stored in traps such as magneto-optical traps (MOTs), pure magnetic traps, and far-off-resonance traps. As discussed in \cite{Weiner1999}, collision studies in such traps are to a certain extent hampered by the lack of a fixed collision axis, masking anisotropic scattering. A few experiments have realized collider-like collision geometries with ultra-cold atoms and demonstrated the power of angular resolved detection of scattered atoms \cite{Legere1998,thomas2004,Ch2004,Mellish2007}. However, these approaches have been severely limited in versatility. For example, the magnetic collider experiments \cite{thomas2004,Ch2004,Mellish2007} were restricted to atoms in weak magnetic field seeking quantum states. Moreover, while the observed quantum scattering interference was quite compelling and showed the qualitative behavior expected from theory, the quantative results for scattering parameters that could be extracted were not of high precision.

\begin{figure}[b!]
\centerline{\includegraphics[width=8.4cm]{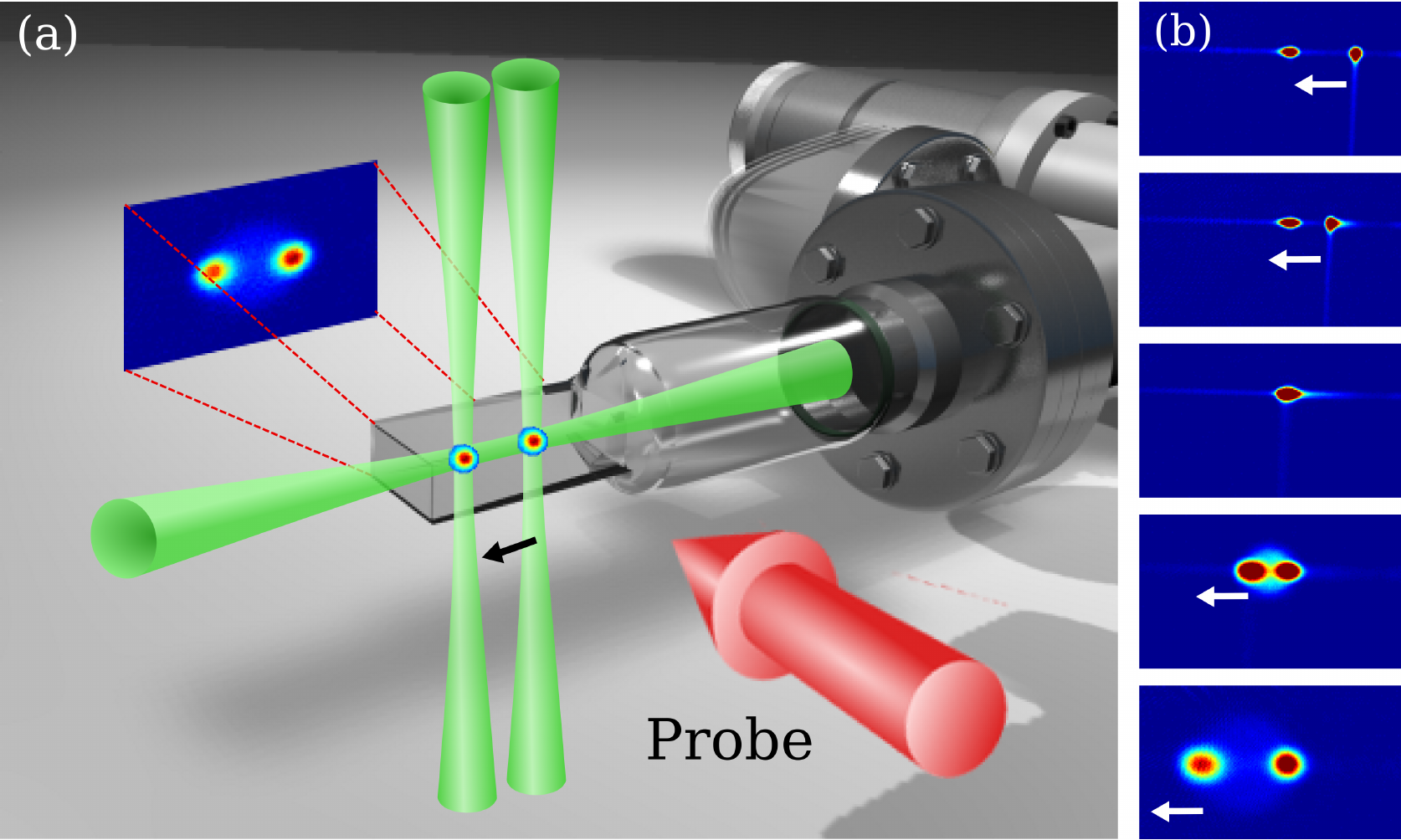}}
\caption{(Color Online) (a) Collider configuration. The right vertical laser beam accelerates atoms towards target atoms in the left crossed-beam trap. The resulting collisional halo is imaged using absorption imaging. (b) Single frame excerpts of a movie (\href{http://www.physics.otago.ac.nz/staff_files/nk/figures/collisionmovie.mov}{Media 1}) showing scattering from clouds in the collider (the horizontal guide beam is extinguished when the atomic clouds overlap so that the clouds and scattered atoms are allowed to expand in free space).}
\label{fig:schematic}
\end{figure}

In this Letter, we demonstrate a collider for ultracold atoms based on optical tweezer beams which guide and accelerate \cite{footnote} the atoms into collisions (see Fig.~\ref{fig:schematic}). Such optical dipole traps for atoms rely on the electric dipole interaction with far off-resonant light \cite{Grimm2000}, and a red-detuned Gaussian laser beam can provide strong radial confinement for atoms. This surpasses the restrictions of magnetic fields, which only provide confinement for certain internal quantum states. In fact, a magnetic field can now be used as a free external tuning parameter for atomic interactions, such that magnetic Fesh\-bach resonances \cite {Chin2010} may be explored in a parameter space spanned by \textit{both} magnetic field strength and collision energy. Moreover, since the acceleration of atoms is dictated directly by dynamical steering of laser beams, the scheme offers exquisite control over the collision energy. In addition to opening up the perspective of laser based colliders as a precision collision metrology tool, this scheme may have the versatility for exploring Efimov physics \cite{Wang2010}, probing strongly-interacting gases \cite{Nishida2011}, and studying atom-dimer and dimer-dimer scattering near Feshbach resonances \cite{Levinsen2011} as recently proposed.

We begin our experimental sequence by cooling a sample of $\rm ^{87}Rb$ atoms in a MOT. After a compressed MOT and molasses stage, atoms are optically pumped into the $\ket{F=2, m_F=2}$ state and captured by a magnetic quadrupole trap mounted on a motorized translation stage. Using the translation stage, the atoms are shuttled to an ultra-high vacuum science cell and transferred into a Ioffe-Pritchard magnetic trap\cite{Ernst1998} where they are cooled to near the Bose-Einstein condensation (BEC) transition using forced radio frequency evaporation. The resulting ultracold sample is transferred into an optical crossed-beam trap. We load the dipole trap by turning on both beams simultaneously over 100 ms and then turning off the magnetic trap over 100 ms. Vertical and horizontal trapping beams with optical powers of 360 mW and 600 mW, focussed to waists of 56 $\mu$m and 60 $\mu$m, respectively, are derived from a 1064 nm Yb fiber laser. Transfer efficiency from the magnetic trap to the optical crossed-dipole trap is 70 $\%$ and we typically load 2\e{6} atoms at 400 nK.

\begin{figure}[t!]
\centerline{\includegraphics[width=8.4cm]{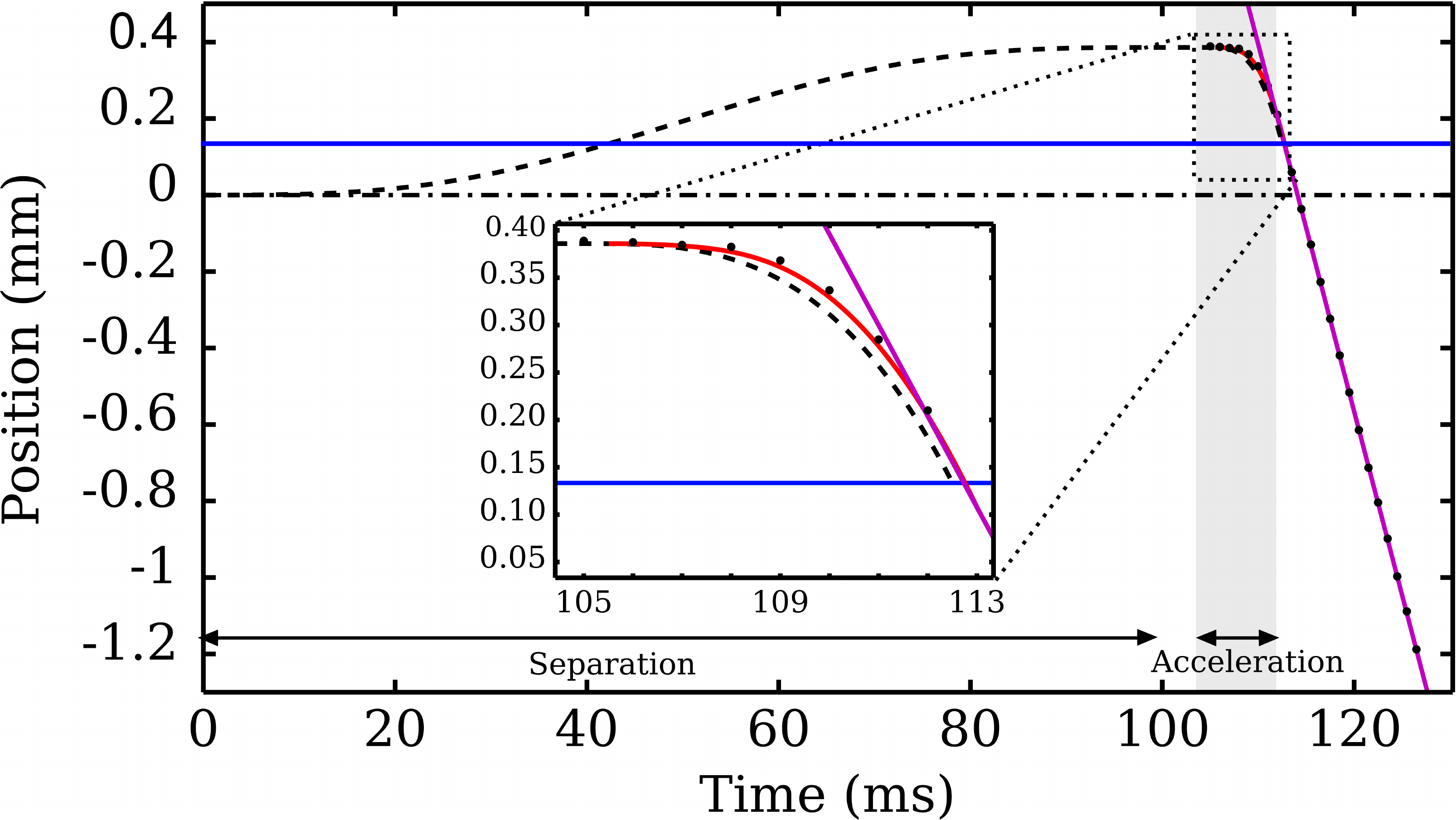}}
\caption{(Color online) Displacement profile (separation and acceleration) for the moving vertical laser beam (dashed line) with respect to the stationary vertical beam (dash-dot line). Points show the experimentally determined positions of the moving cloud after 0.5~ms time of flight (the error bars are smaller than the size of the marker). The blue line indicates the position at which the vertical beams are switched off and the purple line is a linear fit to the cloud position with increasing time of flight. The red line shows the predicted cloud position during the acceleration phase.}
\label{fig:sweep}
\end{figure}

To facilitate a collision, we first split the dipole trapped cloud into two. We achieve this by passing the vertical trapping light through an acousto-optic modulator (AOM) driven by two independently controlled radio frequency sources. This gives rise to two independent crossed dipole traps along the axis of the horizontal trapping beam, which acts as a transport channel. Figure~\ref{fig:schematic}(a) outlines the geometry. The optical layout used for splitting the Gaussian beam into two is similar to \cite{Shin2004,Zawadzki2010}. The cloud splitting must be executed carefully to minimize heating and the beam power in each of the double-well beams must be balanced to ensure that atoms are evenly distributed between the two wells, as this maximizes the number of collision events. We split the cloud using a trapezoidal acceleration profile in time. This reduces sudden changes in acceleration (jerk) which may excite oscillations. Figure~\ref{fig:sweep} shows the resulting S-shaped \cite{Meckl1998} beam position profile in time for the split (first 100~ms). In the demonstration reported here, only one well is displaced; the other (containing the target cloud) remains at its original position. The moving cloud is typically 150 nK hotter than the target cloud. The moving laser beam power varies slightly depending on the AOM driving frequency, but remains within 17 $\%$ of 600 mW during the acceleration phase; the maximum beam separation of 386 $\mu$m is limited by the bandwidth of the AOM, which is not an optimized deflector device. A trap depth of 13 $\mu$K characterizes each well.

After the completion of the split, we hold the atoms for 5 ms before the displaced cloud is accelerated back towards the target atoms held in the stationary crossed-beam trap. The acceleration of the moving beam increases linearly with time at a rate that achieves the desired final velocity at a beam separation of 130 $\mu$m. At this separation, the two vertical beams have no overlap and are both extinguished so that the accelerated cloud can move ballistically along the horizontal guide beam. The motion profile for the moving laser beam used to accelerate the atoms is shown in Fig.~\ref{fig:sweep}. Figure~\ref{fig:schematic}(b)(\href{http://www.physics.otago.ac.nz/staff_files/nk/figures/collisionmovie.mov}{Media~1}) shows absorption images of atoms during this accelerating phase as well as the subsequent scattering on impact, visible as a halo. With the main focus of this Letter resting on the proof of concept for an optical collider, we shall not divert into a further analysis of the scattering halo here; for the energies and conditions encountered, the scattering is dominated by $s$-wave events (i.e. an isotropic distribution).

To demonstrate the high degree of control the optical collider implementation offers, Fig.~\ref{fig:sweep} presents experimentally measured horizontal positions of moving clouds at consecutive time points during and after the acceleration phase. For this data series, clouds were imaged after 0.5~ms time of flight (all beams are turned off) during the acceleration phase in order to avoid resonant frequency shifts when imaging the cloud. The measured positions are in excellent agreement with the expected positions (solid red line) when taking into account this time of flight and that a cloud in the accelerating beam has an equilibrium point offset from the beam center.

\begin{figure}[t!]
\centerline{\includegraphics[width=0.8\columnwidth]{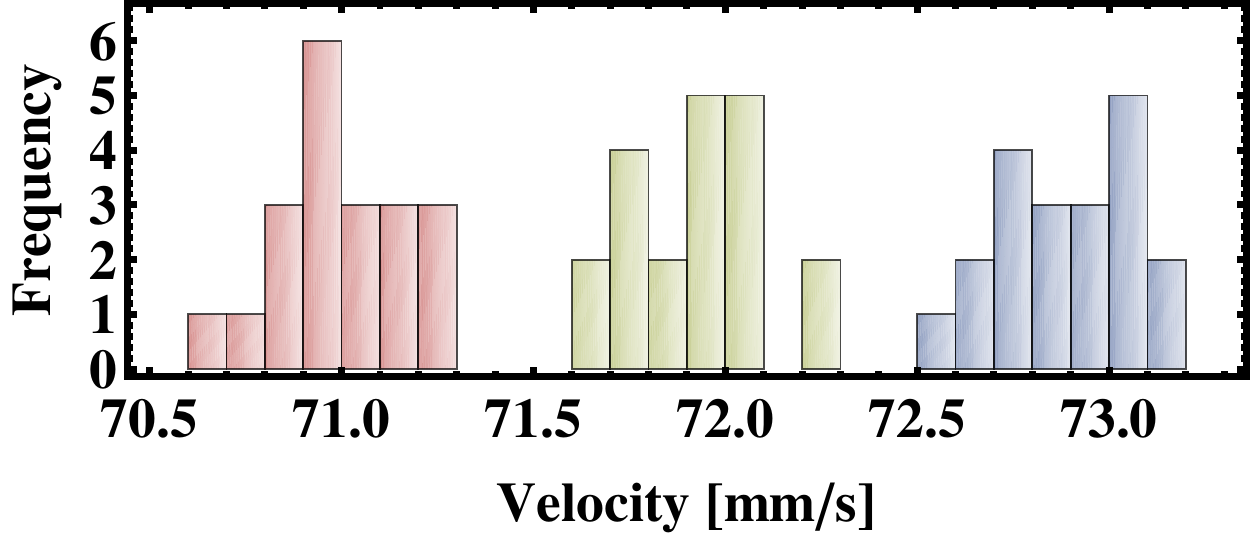}}
\caption{(Color online) Measured cloud velocities when accelerating with three different jerks (red, green, blue).}
\label{fig:hist}
\end{figure}

To further investigate the precision of our collider, we perform multiple experimental realizations of different closely spaced final velocities of the moving cloud. We determine the velocity which was achieved during a given acceleration phase from the measured horizontal positions after 10~ms time of flight for the moving atoms. Figure \ref{fig:hist} shows histograms of measured velocities for 20 realizations at three different set final velocities for the accelerating laser beam. We can clearly reproduce and resolve final velocities within 1~mm/s. Indeed, this discrimination is presently limited by the resolution of our imaging system.

\begin{figure}[t!]
\centerline{\includegraphics[width=8cm]{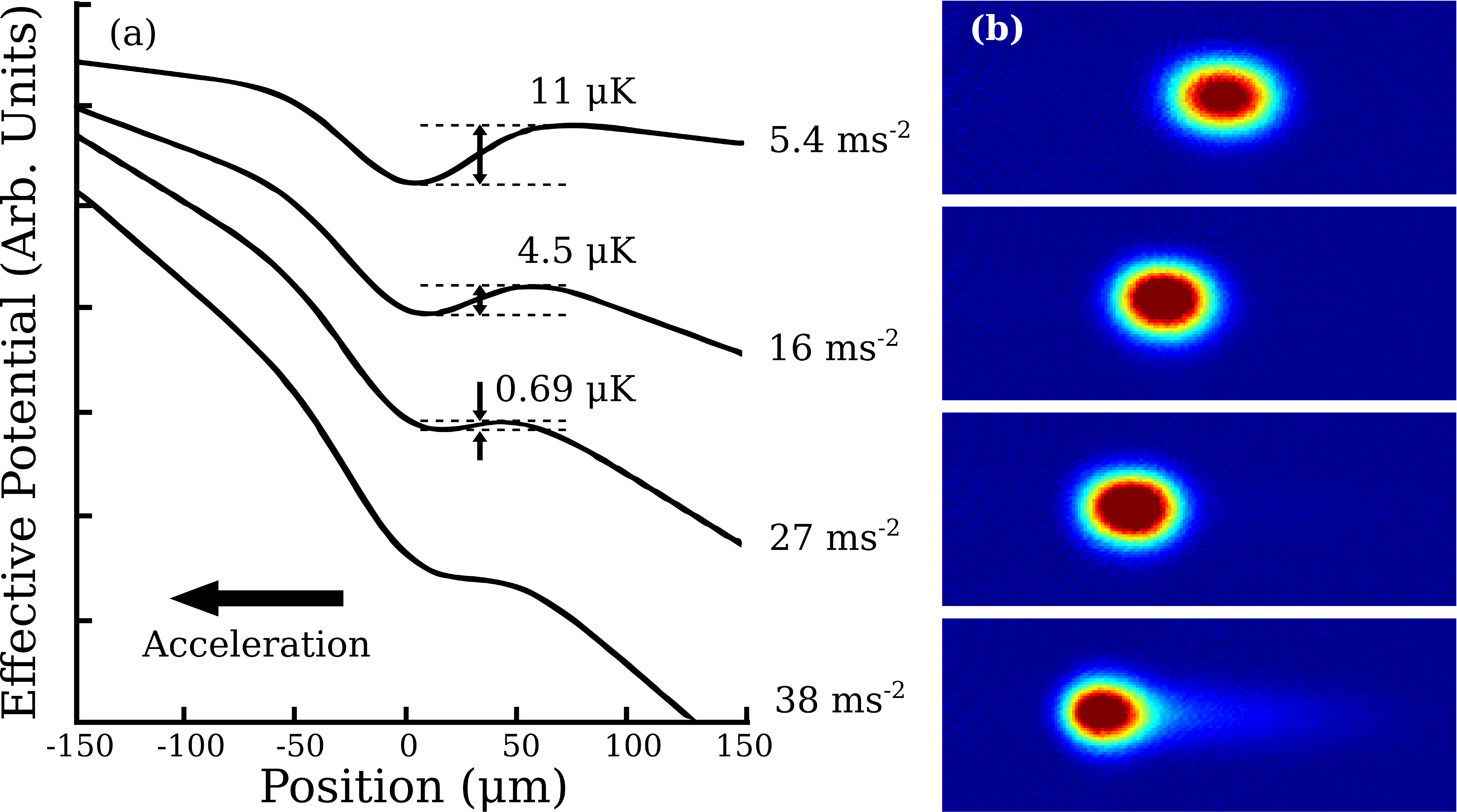}}
\caption{(Color online) (a) Effective potential in a reference frame following a Gaussian laser beam at four different accelerations and (b) corresponding absorption images of clouds (7 ms time of flight) after ramping to these accelerations. The potentials are vertically offset for clarity. Cloud temperature is 250 nK.}
\label{fig:acceleration}
\end{figure}

An important figure of merit for any collider is the maximum energy which particles can be accelerated to. It can be shown that a moving beam of trap depth $U_0$ and beam waist $w_0$ can only support particles up to a maximum acceleration of
\begin{equation}\label{eqmaxax}
a_{\rm max}=\frac{2U_0}{w_0m}\exp(-1/2),
\end{equation}
where $m$ is the particle mass. This can be illustrated via the deformation of the effective potential experienced by a particle in an accelerating reference frame traveling with the laser beam [see Fig.~\ref{fig:acceleration}(a)]. For the laser beam parameters reported here, we estimate $a_{\rm max}=27$~$\rm m/s^2$. This agrees well with our experimental findings [Fig.~\ref{fig:acceleration}(b)], showing that atoms are lost out of the trap if $a_{\rm max}$ is exceeded during the acceleration phase. For an acceleration increasing linearly in time from zero to $a_{\rm max}$ applied over a distance $d_{\rm sep}$, the final velocity is
\begin{equation}\label{eqvfin}
v_f=\sqrt{3a_{\rm max}d_{\rm sep}/2}.
\end{equation}
Using the scheme and parameters outlined in the present Letter, we obtain $v_f=9.8$~cm/s which corresponds to a collision energy (center of mass frame) of $\sim50$~$\rm\mu K$ as measured in units of the Boltzmann constant. This corresponds well with the maximum energies achieved in our present setup. Equation (\ref{eqvfin}) shows the collider energy grows linearly with $d_{\rm sep}$. We are presently upgrading our collider with a dedicated acousto-optic deflector device, which, in conjunction with counter-accelerating the two clouds and increasing the optical powers of the tweezer beams (thereby increasing $U_0$), should take us into the milliKelvin regime.

In conclusion, we have characterized an ultracold atom collider based on moving laser tweezers that offers eminent control in manipulating and accelerating atomic clouds. In conjunction with recent progress on direct simulation Monte Carlo methods applied to the Boltzmann equation in the quantum collision regime \cite{Wade2011}, we expect our laser based accelerator scheme to develop into a high precision tool for collision metrology. In addition to Rb, our setup comprises K as a species and future directions will include inter-species collisions and studies of fermionic $\rm ^{40}K$-$\rm^{40}K$ scattering at the $p$-wave threshold.

We note that while completing this manuscript, angularly resolved, non-trivial scattering was reported for ``dressed" BECs colliding at energies below 1~$\mu$K \cite{Williams2011}.

This work was supported by FRST
Contract No.~NERF-UOOX0703.
\bibliographystyle{ol}

\begin{thebibliography}{10}
\newcommand{\enquote}[1]{``#1''}

\bibitem{Taylor2006}
J.~R. Taylor, \emph{Scattering Theory} (Dover, 2006).

\bibitem{Weiner1999}
J.~Weiner, V.~S. Bagnato, S.~Zilio, and P.~S. Julienne, Rev. Mod. Phys.
  \textbf{71}, 1 (1999).

\bibitem{Legere1998}
R.~Legere and K.~Gibble, Phys. Rev. Lett. \textbf{81}, 5780 (1998).

\bibitem{thomas2004}
N.~R. Thomas, N.~Kj{\ae}rgaard, P.~S. Julienne, and A.~C. Wilson, Phys. Rev.
  Lett. \textbf{93}, 173201 (2004).

\bibitem{Ch2004}
Ch.~Buggle, J.~L\'{e}onard, W.~von Klitzing, and J.~T.~M. Walraven, Phys. Rev.
  Lett. \textbf{93}, 173202 (2004).

\bibitem{Mellish2007}
A.~S. Mellish, N.~Kj{\ae}rgaard, P.~S. Julienne, and A.~C. Wilson, Phys. Rev. A
  \textbf{75}, 020701 (2007).

\bibitem{footnote}
Outside the context of colliders, laser acceleration of cold atoms has been demonstrated using far detuned ``moving standing waves'', see, e.g., E.~Peik, M.~Ben~Dahan, I.~Bouchoule, Y. Castin, and C.~Salomon, Appl. Phys. B \textbf{65}, 685 (1997).

\bibitem{Grimm2000}
R.~Grimm, M.~Weidem\"{u}ller, and Y.~B. Ovchinnikov, Adv. At. Mol. Opt. Phys.
  \textbf{42}, 95 (2000).

\bibitem{Chin2010}
C.~Chin, R.~Grimm, P.~Julienne, and E.~Tiesinga, Rev. Mod. Phys. \textbf{82},
  1225 (2010).

\bibitem{Wang2010}
Y.~Wang, J.~P. D'Incao, H.~C. N\"{a}gerl, and B.~D. Esry, Phys. Rev. Lett.
  \textbf{104}, 113201 (2010).
\bibitem{Nishida2011}
Y.~Nishida, http://arxiv.org/abs/1110.5926.
\bibitem{Levinsen2011}
J.~Levinsen and D.~S.~Petrov, Eur. Phys. J. D \textbf{65}, 67 (2011).

\bibitem{Ernst1998}
U.~Ernst, A.~Marte, F.~Schreck, J.~Schuster, and G. Rempe, Europhys. Lett.
  \textbf{41}, 1 (1998).

\bibitem{Shin2004}
Y.~Shin, M.~Saba, T.~A. Pasquini, W.~Ketterle, D.~E. Pritchard, and A.~E.
  Leanhardt, Phys. Rev. Lett. \textbf{92}, 050405 (2004).

\bibitem{Zawadzki2010}
M.~E. Zawadzki, P.~F. Griffin, E.~Riis, and A.~S. Arnold, Phys. Rev. A
  \textbf{81}, 043608 (2010).

\bibitem{Meckl1998}
P.~Meckl, P.~Arestides, and M.~Woods, in \textit{Proceedings of the American Control
  Conference} vol.~5, pp.~2627-2631 ({1998}).

\bibitem{Wade2011}
A.~C.~J. Wade, D.~Baillie, and P.~B. Blakie, Phys. Rev. A \textbf{84}, 023612
  (2011).

\bibitem{Williams2011}
R.~A. Williams, L.~J. LeBlanc, K.~Jim\'{e}nez-Garc\'{\i}a, M.~C. Beeler, A.~R.
  Perry, W.~D. Phillips, and I.~B. Spielman, Science \textbf{335}, 314 (2012).

\end{thebibliography}

\end{document}